\documentclass[aps,prl,superscriptaddress,preprintnumbers,reprint,nofootinbib,showpacs]{revtex4-1}
\usepackage{graphicx}
\usepackage{epsfig}
\usepackage{amssymb}

\def\als{\alpha_s} 
\def\MS{\overline{\rm MS}}
\newcommand{\be}{\begin{equation}}
\newcommand{\ee}{\end{equation}}
\newcommand{\bea}{\begin{eqnarray}}
\newcommand{\eea}{\end{eqnarray}}

\begin{document}

\title{Determination of $\alpha_s$ from the QCD static energy}
\author{Alexei Bazavov}
\affiliation{Physics Department, Brookhaven National Laboratory,
  Upton, NY 11973, USA}
\author{Nora Brambilla}
\affiliation{Physik Department, Technische Universit\"at M\"unchen, D-85748 Garching, Germany}
\author{Xavier \surname{Garcia i Tormo}}
\affiliation{Albert Einstein Center for Fundamental Physics. Institut f\"ur Theoretische Physik, Universit\"at Bern,
  Sidlerstrasse 5, CH-3012 Bern, Switzerland}
\author{P\'eter Petreczky}
\affiliation{Physics Department, Brookhaven National Laboratory,
  Upton, NY 11973, USA}
\author{Joan Soto}
\affiliation{Departament d'Estructura i Constituents de la Mat\`eria and Institut de Ci\`encies del Cosmos, 
Universitat de Barcelona, Diagonal 647, E-08028 Barcelona, Catalonia, Spain}
\author{Antonio Vairo}
\affiliation{Physik Department, Technische Universit\"at M\"unchen, D-85748 Garching, Germany}

\date{\today}

\preprint{TUM-EFT 31/12,~   UB-ECM-PF 11/71, ~ICCUB-12-122}

\begin{abstract}
We compare lattice data for the short-distance part of the static
energy in $2+1$ flavor quantum chromodynamics (QCD) with perturbative
calculations, up to next-to-next-to-next-to leading-logarithmic accuracy. We show that perturbation theory describes very well the
lattice data at
short distances, and exploit this fact to obtain a determination of
the product of the lattice scale $r_0$ with the QCD scale
$\Lambda_{\MS}$. With the input of the value of $r_0$, this provides a
determination of the strong coupling $\alpha_s$ at the typical
distance scale of the lattice data. We obtain
$\alpha_s\left(1.5{\rm GeV}\right)=0.326\pm0.019$,
which provides a novel determination of $\alpha_s$ with three-loop accuracy (including resummation
of the leading ultrasoft logarithms), and constitutes one of the few
low-energy determination of $\alpha_s$ available. When this value is evolved to the $Z$-mass scale $M_Z$, it corresponds to $\alpha_s\left(M_Z\right)=0.1156^{+0.0021}_{-0.0022}$.
\end{abstract}

\pacs{12.38.Aw, 12.38.Bx, 12.38.Gc}

\maketitle

The static energy in quantum chromodynamics (QCD), i.e. the energy
between a static quark and a static antiquark separated by a distance $r$, is a basic object to
understand the behavior of the theory \cite{Wilson:1974sk} and
constitutes a fundamental ingredient in the description of many
physical processes \cite{Brambilla:2010cs}. The
short-distance part of the static energy can be computed using perturbative techniques, and
it is nowadays known at next-to-next-to-next-to leading-logarithmic (N$^3$LL) accuracy,
i.e. including terms up to order $\als^{4+n}\ln^n\als$ with $n\ge
0$
\cite{Brambilla:2010pp,Pineda:2011db,Brambilla:2009bi,Smirnov:2009fh,Anzai:2009tm,Smirnov:2008pn,Brambilla:2006wp}
($\ln\als$ terms appear due to virtual
emissions of ultrasoft gluons, which can change the color state of the
quark-antiquark pair \cite{Appelquist:1977es,Brambilla:1999qa}). It can also be computed on the lattice, and the comparison of the
two approaches tests our ability to describe the short-distance regime
of QCD, besides providing information on the region of validity of the
perturbative weak-coupling approach~\cite{Billoire:1981ny}. A comparison of the static energy
at N$^3$LL accuracy with quenched lattice data \cite{Necco:2001xg} was
presented in Ref.~\cite{Brambilla:2010pp}. Here we present
lattice data for the short-distance part of the static energy in $2+1$ flavor QCD and compare it
with the perturbative calculation up to N$^3$LL accuracy. This allows us
to determine the strong coupling $\alpha_s$ at three-loop accuracy (including resummation
of the leading ultrasoft logarithms), in a way which is largely
independent from the other determinations that currently enter in the
world average \cite{Bethke:2009jm}. The natural scale
where our determination is performed corresponds to the inverse of the
typical distance where we have lattice data, i.e. around
$1.5$ GeV. Therefore, our analysis provides a determination
of $\als$ at a scale smaller than those entering the current world
average \cite{Bethke:2009jm}, and constitutes in this way an important
ingredient to further test asymptotic freedom in QCD.

\begin{figure}
\centering
\includegraphics[width=8.6cm]{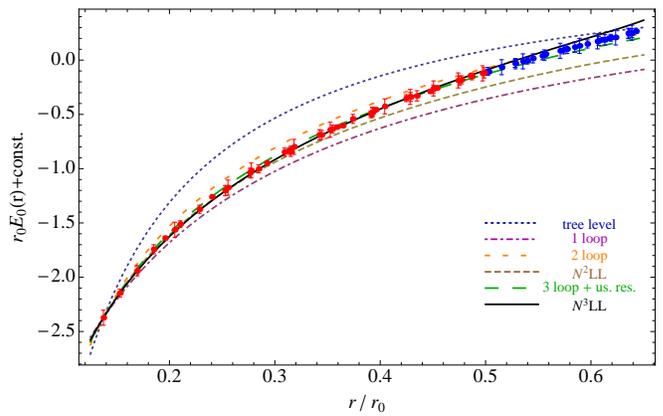}
\caption{Comparison of the singlet static energy with lattice data
  (red -lighter- points). [The comparison (and all the analysis in the text) is
  done for $r<0.5r_0\simeq0.234$ fm, which is the region where
  perturbation theory is reliable. The (blue -darker-) points and curves for $r>0.5r_0$
  are shown just for illustration]. The dotted blue
  curve is at tree level, the dot-dashed magenta curve is at one loop,
  the long-spaced-dashed orange curve is at two loops, the
  dashed brown curve is at N$^2$LL accuracy, the long-dashed green curve is at
  three loops plus leading ultrasoft logarithmic resummation, and the solid black curve is at N$^3$LL accuracy. $r_0\Lambda_{\MS}=0.70$ was used
  in all the curves. The additive constant in the perturbative
  expression for the static energy is
  taken such that each curve coincides with the lattice data point at
  the shortest distance.
}\label{fig:enrhomid}
\end{figure}

The static energy has been calculated on the lattice in $2+1$ flavor QCD 
using a combination of tree-level improved gauge action and highly-improved staggered 
quark (HISQ) action \cite{Follana:2006rc} in Ref.~\cite{Bazavov:2011nk}. The strange-quark mass $m_s$ 
was fixed to its physical value, while the light-quark masses were
chosen to be $m_l=m_s/20$. These correspond to the pion mass of about 
$160$ MeV in the continuum limit, which is very close to the physical value.
The calculation 
of the static energy was performed in a wide range of gauge
couplings $5.9\le \beta \equiv 10/g^2 \le 7.28$.
At each value of the gauge coupling we calculate the scale parameters $r_0$ and $r_1$ defined
in terms of the static energy $E_0(r)$ as follows \cite{Sommer,milc04}
\begin{equation}
r^2 \frac{d E_0(r)}{d r}|_{r=r_0}=1.65,~~~r^2 \frac{d E_0(r)}{d r}|_{r=r_1}=1.
\end{equation}
The values of $r_0$ and $r_1$ were given in Ref. \cite{Bazavov:2011nk} for each $\beta$.
The above range of the gauge couplings corresponds to lattice spacing 
$1.909/r_0 \le a^{-1} \le 6.991/r_0$. Using the most recent value $r_0=0.468\pm0.004$ fm
\cite{Bazavov:2011nk} we get $0.805\,{\rm GeV}<a^{-1}< 2.947\,{\rm GeV}$. 
Thus we can study the static energy down to distances $r=0.14r_0$ or
$r\simeq 0.065$ fm. 
For the comparison with perturbation theory the most relevant data set is the one 
corresponding to lattice gauge coupling $\beta = 6.664, 6.740, 6.800, 6.880, 6.950, 7.030 , 7.150,
7.280$, which is what we will use here.  The static energy can be calculated in units of $r_0$ or $r_1$. 
Since the static energy has an additive ultraviolet renormalization we need to normalize the results 
calculated at different lattice spacings to a common value at a
certain distance. We fix the static energy in units of $r_0$ to 0.954 at
$r = r_0$ \cite{Bazavov:2011nk}.
At distances comparable to the lattice spacing the static energy suffers from lattice
artifacts. To correct for these artifacts we use tree level improvement. 
From the lattice Coulomb potential 
\begin{equation}
C_L(r)=\int\frac{d^3 k}{(2 \pi)^3} D_{00}(k_0=0,\vec{k}) e^{i \vec{k} \vec{r}},
\end{equation}
we can define the improved distance $r_I=(4 \pi C_L(r))^{-1}$ for each
separation $r$. Here $D_{00}$ is the tree level gluon propagator for the
$a^2$ improved gauge action. The tree level improvement amounts to replacing
$r$ by $r_I$ \cite{Necco:2001xg}. Alternatively following Ref. \cite{milc04,ukqcd}
we fit the lattice data at short distances to
the form $const-a/r +\sigma r+a' (1/r-1/r_I)$ and subtract the last term from
the lattice data. We have found that both methods of correcting for lattice
artifacts lead to the same results within errors of the calculations. Furthermore,
the static energies calculated for different lattice spacings agree well with
each other after the removal of lattice artifacts.  
The corrected  lattice data obtained for several lattice spacings are
shown in Fig.~\ref{fig:enrhomid} as the points. All the lattice data seem
to lie on a single curve even at short distances, indicating that the above
procedure of removing the lattice artifacts works.

As mentioned before, the static energy is known at N$^3$LL accuracy in
perturbation theory. 
Detailed expressions
for $E_0$ were given in Ref.~\cite{Brambilla:2010pp} (and references
therein) and will not be reproduced here. For our present analysis, it
is only important to recall that:
(i) In order to obtain a well
behaved perturbative series, it is necessary to implement a scheme
that cancels the leading renormalon singularity
\cite{Beneke:1998rk}. This kind of schemes introduce
dependence on a dimensional scale in the problem, which we
denote as $\rho$. In particular, we implement the renormalon
cancellation according to the scheme described in
Ref.~\cite{Pineda:2001zq}. Then, the natural value for the scale $\rho$
corresponds to the center of the range where we have lattice data (ii)
At N$^3$LL accuracy the perturbative expression depends on an
additional constant (which was not present at lower levels of
accuracy). This is due to the structure of the renormalization group
equations at that order. We call this constant $K_2$. It should
satisfy the power counting condition $|K_2|\sim\Lambda_{\MS}$ (where
$\Lambda_{\MS}$ is the QCD scale in the $\MS$ scheme), but apart from
that it is unconstrained.

We can now compare the perturbative results for the static energy with the lattice data. This comparison goes along the same lines as the
quenched case in Ref.~\cite{Brambilla:2010pp}, except that now we use
$n_f=3$ everywhere ($n_f$ is the number of light flavors); we
  also include finite strange mass effects at one loop, although they turn
  out to be negligible. We use the maximum known accuracy (four loop)
for the running of $\alpha_s$ everywhere (as opposed to changing the
accuracy for the running depending on the order we are working at), since, for the $n_f=3$ case, the hierarchy of
scales underlying the perturbative calculation would not be well
satisfied with the running of $\alpha_s$ at one loop. The perturbative
expressions depend on the value of the quantity $r_0\Lambda_{\MS}$,
and we will use the lattice data to determine it. That is, we search for the
range of $r_0\Lambda_{\MS}$ that is allowed by lattice data, taking
into account all the uncertainties involved. Then, using the value for $r_0$
determined in Ref.~\cite{Bazavov:2011nk} we can obtain a determination
of $\alpha_s(M_Z)$ ($M_Z$ is the $Z$-boson mass).

Under the assumption that perturbation theory by itself (after
canceling the leading renormalon) is enough to accurately describe
the lattice data in the range of distances we are considering (i.e. $r<0.5r_0$), a procedure to extract $r_0\Lambda_{\MS}$ from the comparison of the
perturbative expressions for the static energy with lattice data was
devised in Ref.~\cite{Brambilla:2010pp}. We will proceed in an
analogous way here. The procedure exploits the fact that any value
of $\rho$ (around its natural value at the center of the range for
which we have lattice data, i.e. $\rho=3.14r_0^{-1}$) cancels the renormalon and is
therefore allowed. Following Ref.~\cite{Brambilla:2010pp}, we search for a set of $\rho$ values
which are optimal for the determination of $r_0\Lambda_{\MS}$. The
procedure to do that consists of the following steps:
\begin{enumerate}
\item We vary $\rho$ by $\pm 25\%$ around its natural value
  $\rho=3.14r_0^{-1}$, that is from $\rho=2.36r_0^{-1}$ to
  $\rho=3.93r_0^{-1}$.
\item For each value of $\rho$ and at each order in the perturbative
  expansion of the static energy, we perform a fit to the lattice
  data (that is we do fits at tree level, one loop, two loops, and three loops;
  in the last two cases with and without ultrasoft logarithmic resummation). The parameter in each of these fits is $r_0\Lambda_{\MS}$.
\item We select those $\rho$ values for which the reduced $\chi^2$ of
  the fits decreases when increasing the number of loops of the perturbative calculation.
\end{enumerate}
For the analysis, we use the fits from tree level to three loop plus
leading ultrasoft logarithmic resummation accuracy. We could also use the fits at N$^3$LL accuracy (i.e. three loops plus
sub-leading ultrasoft logarithmic resummation), which would involve
the additional constant $K_2$ (that would also need to be fitted to
the data, i.e. the fits involve one additional parameter at this order). We find that, with the present
lattice data, the $\chi^2$ as a
function of $r_0\Lambda_{\MS}$  is very flat in this case, and we
cannot improve our extraction of $r_0\Lambda_{\MS}$ by including
the fits at N$^3$LL accuracy in the analysis. We interpret this as the unquenched lattice data not being
accurate enough to be sensitive to subleading ultrasoft logarithms
(unlike the quenched one used in \cite{Brambilla:2010pp}), a fact that
leaves room for future improvements. Therefore, we take the numbers at
three loop plus leading ultrasoft logarithmic resummation accuracy as
our best result, and consider, at this order, the set of fitted values
of $r_0\Lambda_{\MS}$ for the $\rho$ range obtained after step 3 above
(we denote these values by $x_i$). We assign a weight to each $x_i$, given by the inverse of the
reduced $\chi^2$ of the fit. We take the weighted average of the $x_i$ as our central value for the
determination of $r_0\Lambda_{\MS}$. To estimate the error that we should associate
to this number, we consider the weighted standard deviation of this
set of values, and the difference with the
weighted average computed using the result at the previous
perturbative order (with the corresponding $\rho$ range that one
obtains at that order; for illustration, we show the results for
$r_0\Lambda_{\MS} $ obtained at different levels of accuracy in
Tab.~\ref{tab:Lam}). We obtain $r_0\Lambda_{\MS}=0.7024\pm0.0011\pm0.0665=0.70\pm0.07$,
where the first error is due to the weighted standard deviation, the
second to the difference with the two-loop result, and we summed the
two errors linearly on the right-hand side of the equation. It is
important to point out that the error
assigned to the result must account for the uncertainties associated
to the neglected higher-order terms in the perturbative expansion of
the static energy; in that sense, assigning the difference with the
result at the previous order as an error (as we do) is a quite
conservative estimate. 
\begin{table}
\begin{tabular}{c|c|c}
Accuracy & $r_0\Lambda_{\MS}$ &  $\frac{r_0}{r_1}\, r_1\Lambda_{\MS}$\\
\hline tree level & $0.395$ & $0.397$ \\
1 loop & $0.848$ & $0.862$ \\
2 loop & $0.636$ & $0.654$ \\
N$^2$LL & $0.756$ & $0.783$ \\
3 loop & $0.690$ & $0.701$ \\
3 loop + us. res. & $0.702$ & $0.715$ \\
\end{tabular}
\caption{Values of $r_0\Lambda_{\MS}$ obtained at different levels of
  accuracy. The second column shows the results obtained using the
  static energy normalized in units of the scale $r_0$. The third
  column shows the results obtained with the
  static energy normalized in units of the scale $r_1$ and then (for
  easier comparison) transformed to $r_0$ units, using the factor
  $r_0/r_1=1.508\pm0.005$ \cite{Bazavov:2011nk}. ``N$^2$LL'' stands for next-to-next-to leading-logarithmic
  (i.e. two loop plus leading ultrasoft logarithmic resummation) and ``3
  loop + us. res.''  stands for three loop plus leading ultrasoft logarithmic resummation.}\label{tab:Lam}
\end{table}
Note that, starting at the two-loop level, one can decide whether to
perform resummation of the ultrasoft logarithms or not; when assigning
the error, we take whichever difference is larger. We also mention that there is an error associated
to each of the $x_i$ coming from the fit to the lattice data, but the error that this induces in the
average can be neglected. To further assess the systematic errors stemming from our
procedure, we have redone the analysis using $p$-value weights,
obtaining $r_0\Lambda_{\MS}=0.7022\pm0.0011\pm0.0628=0.70\pm0.06$, and
using constant weights, obtaining
$r_0\Lambda_{\MS}=0.7022\pm0.0011\pm0.0666=0.70\pm0.07$, which are both
compatible with the previous analysis. In our final result we quote an error that covers the whole range spanned by the three
analyses. As an additional cross-check of the result, we
have redone the analysis with the static energy normalized in units
of the scale $r_1=0.3106\pm 0.0020$ fm (rather than $r_0$); these numbers
are presented in the third column of Tab.~\ref{tab:Lam}, and are
consistent with our previous results. Finally, we point out that in some cases the $\chi^2$ as a
function of $r_0\Lambda_{\MS}$ at next-to-next-to leading-logarithmic
(N$^2$LL) accuracy develops a second local minimum for larger values of $r_0\Lambda_{\MS}$. To
discern which minimum should be taken as the physical result when this happens, we have
redone the fits using smaller $r$ ranges. We found that the position
of the second minimum is not stable, while the position of the first
one is. Furthermore the minima of the $\chi^2$ from lower orders in
perturbation theory come closer to the first of the two minima at
N$^2$LL, when decreasing the $r$ range we use. In view of the above,
when a second minimum develops, we keep the first one, which is stable and
preferred by lower perturbative orders.

According to the discussion in the previous paragraph, our final result reads
\begin{equation}\label{eq:Lambda}
r_0\Lambda_{\MS}=0.70\pm0.07,
\end{equation}
which using $r_0=0.468\pm0.004$ fm \cite{Bazavov:2011nk} gives
\begin{equation}\label{eq:as1p5}
\alpha_s\left(\rho=1.5{\rm GeV},n_f=3\right)=0.326\pm0.019,
\end{equation}
the uncertainty in $r_0$ is negligible in the final error above. When we evolve Eq.~(\ref{eq:as1p5}) to the scale $M_Z$ we obtain
\begin{equation}\label{eq:asMZ}
\alpha_s\left(M_Z,n_f=5\right)=0.1156^{+0.0021}_{-0.0022},
\end{equation}
where we have used the \verb|Mathematica| package \verb|RunDec| \cite{Chetyrkin:2000yt} to obtain
the above number (4 loop running, with the charm quark mass equal to
1.6 GeV and the bottom quark mass equal to 4.7 GeV). We mention that
the final result employing the static energy normalized in units of $r_1$
is $\alpha_s(M_Z)=0.1160^{+0.0021}_{-0.0022}$, which is compatible with
our result in Eq.~(\ref{eq:asMZ}) and further shows its robustness. Figure
\ref{fig:enrhomid} shows a comparison of the perturbative expressions
for the static energy with lattice data using our result
$r_0\Lambda_{\MS}=0.70$ in Eq.~(\ref{eq:Lambda}) ($\rho$ is set at the natural
value, $\rho=3.14r_0^{-1}$). We can see that the
perturbative series converges, approaches the lattice data, and
reproduces it very well at N$^3$LL accuracy (the constant $K_2$ that
appears at N$^3$LL is fixed by a fit to the lattice data, which gives
$r_0K_2=-2.3$, fulfilling the power counting condition)
. Note also
that our results for $\alpha_s$ are not sensitive to the specific value of $r$ that
we consider as the upper limit where perturbation theory is reliable, as it is manifest from Fig.~\ref{fig:enrhomid}.

Our result in Eq.~(\ref{eq:asMZ}) constitutes a novel determination of
$\alpha_s$ (since it is largely
independent of the other available determinations), that stems from a perturbative calculation of the QCD
static energy at three loop plus leading ultrasoft logarithmic
resummation accuracy. With respect to the other determinations currently entering
    the world average it represents the one at lowest energy. The
    lowest-energy determination so far was that from the $\tau$
    system, performed at $m_{\tau}=1.78$ GeV. Our result is therefore
    an important new ingredient to test the running of $\als$. 

Other recent
determinations of $\alpha_s$, that also employ comparisons with
lattice data, include Refs.~\cite{Davies:2008sw,McNeile:2010ji} where several
observables related to Wilson loops (but not the static energy) are
used, Refs.~\cite{Allison:2008xk,McNeile:2010ji} which employ moments
of heavy quark correlators, Ref.~\cite{Shintani:2010ph} that uses the
  vacuum polarization function, Ref.~\cite{Aoki:2009tf} which uses the so-called
Schr\"odinger functional scheme (albeit employing rather high pion
masses), and Ref.~\cite{Blossier:2012ef} that employs the ghost-gluon coupling; they
deliver numbers that are mostly compatible with our
result, although with central values a bit higher than ours. We also mention that comparisons of perturbative calculations for the
static energy with lattice data in QCD with $n_f=2$ flavors have been
presented recently in Refs.~\cite{Jansen:2011vv,Knechtli:2011pz}.

Let us also point out that the comparison of the perturbative result
with lattice data, shown in Fig.~\ref{fig:enrhomid}, is interesting in
itself, since the static energy constitutes a basic ingredient in the
description of many physical processes~\cite{Brambilla:2010cs}. As an example, it is relevant for the study of quarkonium production in
heavy-ion collisions: A direct lattice calculation of the quarkonium
spectral functions is known to be difficult
\cite{Petreczky:2010yn}. The only viable option could be calculating quarkonium spectral
functions within an effective field theory framework, as potential
Non-Relativistic QCD (pNRQCD) provides \cite{Pineda:1997bj,Brambilla:1999xf}. To access the validity
of a weak coupling pNRQCD approach at non-zero temperature, one
eventually will have to compare weak-coupling calculations of the
static quark-antiquark correlators with the corresponding lattice
calculations. The comparison of the static energy at zero temperature to
the perturbative results, in 2+1 flavor QCD with physical quark masses
(as we have provided), is an important first step in this direction.

\begin{figure}
\centering
\includegraphics[width=8.6cm]{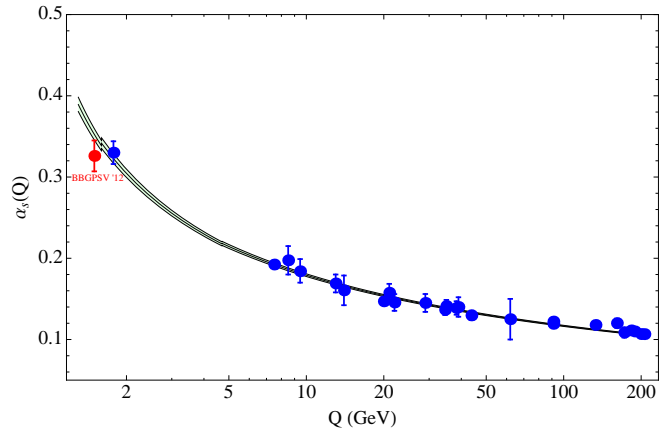}
\caption{Determinations of $\alpha_s$ that enter in the world average
  \cite{Bethke:2009jm} (blue -darker- points) compared with our result
  (most-left red -lighter-
  point), as a function of the energy scale $Q$. The band is the
  world-average value of $\alpha_s(M_Z)$, evolved with 4-loop accuracy.}\label{fig:compas}
\end{figure}

In summary, we have shown that perturbation theory (after canceling
the leading renormalon singularity) can describe the short-distance
part of the QCD static energy computed on the lattice, see Fig.~\ref{fig:enrhomid}; this is the
first time that this is done for QCD with $n_f=2+1$ dynamical quarks. Exploiting this fact,
we have obtained the range of $r_0\Lambda_{\MS}$ that is allowed by
lattice data. Using the value of $r_0$ as an additional input, this
provides a determination of $\alpha_s$. We obtained
$\alpha_s\left(1.5{\rm GeV}\right)=0.326\pm0.019$,
which represents one of the few low-energy determinations of $\als$
available and, when evolved to the scale $M_Z$, corresponds to
$\alpha_s\left(M_Z\right)=0.1156^{+0.0021}_{-0.0022}$. A comparison of
our result with the determinations of $\alpha_s$ that currently enter in the
world average \cite{Bethke:2009jm} is shown in Fig.~\ref{fig:compas}.

\begin{acknowledgments}
We thank Philippe de Forcrand for pointing out
Ref.~\cite{Billoire:1981ny}. This work was supported in part by U.S. Department of Energy under
Contract No. DE-AC02-98CH10886. N.B. and A.V. acknowledge
financial support from the DFG cluster of excellence ``Origin and
structure of the universe'' (www.universe-cluster.de). J.S. is supported by the CPAN CSD2007-00042 Consolider-Ingenio 2010 program (Spain), the 2009SGR502 CUR grant (Catalonia) and the FPA2010-16963 project (Spain).
\end{acknowledgments}

\end{document}